\pdfoutput=1
\documentclass{PoS}
\usepackage{braket}
\usepackage{amsmath}

\title{The radiative corrections to double-Dalitz decays revisited}

\ShortTitle{The radiative corrections to double-Dalitz decays revisited}

\author{
        \speaker{Pablo Sanchez-Puertas}$^{\ a,b}$\\
        \llap{$^a$}Institut de F\'isica d'Altes Energies (IFAE),
        The Barcelona Institute of Science and Technology (BIST),  
        Universitat Aut\`onoma de Barcelona, E-08193 Bellaterra (Barcelona), Spain,\\
        \llap{$^b$}Faculty of Mathematics and Physics, Institute of Particle and Nuclear Physics,
        Charles University in Prague, V Hole{\v{s}}ovi{\v{c}}k{\'a}ch 2, 18000 Praha 8, Czech Republic
        E-mail: \email{psanchez@ifae.es}}

\abstract{
We revise the radiative corrections to double-Dalitz decays 
($P\to\ell\bar{\ell}\ell'\bar{\ell}'$), completing the full next-to-leading order 
calculation in QED as compared to existing calculations. As a result, we find mild 
differences with respect to previous studies, that might be relevant for extracting 
information about the mesons transition form factors. The latter play an important 
role in determining the hadronic light-by-light contribution to the anomalous magnetic 
moment of the muon. Finally, we outline an ongoing extension of this work to the 
$e^+e^-\to e^+e^- P$ processes.
}

\FullConference{The 9th International workshop on Chiral Dynamics\\
		17-21 September 2018\\
		Durham, NC, USA}

\begin{document}

\section{Introduction}

The transition form factors (TFFs) of pseudoscalar mesons ($P=\pi^0,\eta,\eta'$), commonly denoted as $F_{P\gamma^*\gamma^*}(q_1^2,q_2^2)$, describe the interaction of the former with two---possibly virtual---photons
\begin{equation}
   i\int d^4x \ e^{iq_1\cdot x} \bra{0} T\{ j^{\mu}(x) j^{\nu}(0) \} \ket{P} = 	\epsilon^{\mu\nu\rho\sigma}q_{1\rho}q_{2\sigma} F_{P\gamma^*\gamma^*}(q_1^2,q_2^2).
\end{equation}
They have been an object of interest in the past and present years. From one point of view, such an exclusive process with a single---and a priori the simplest---QCD bound state, represents in principle the cleaner one that can be described in pQCD (see Refs.~\cite{Lepage:1980fj,Lepage:1979zb} for pioneering works). As such, it serves as a probe of our understanding of QCD dynamics and the pseudoscalar meson structure itself. From a different point of view, and more recently, these have regained interest in connection with their role in the anomalous magnetic moment of the muon, $(g-2)_{\mu}$~\cite{Jegerlehner:2009ry,Benayoun:2014tra,Masjuan:2017tvw}, which at the moment shows an interesting $3\sigma$ deviation~\cite{Jegerlehner:2009ry} and for which  new---more precise results---are expected soon from the new experiment at Fermilab~\cite{LeeRoberts:2011zz} and also, in the future, from J-PARC~\cite{Mibe:2010zz}.

A key ingredient here is the description of these form factors at the low energies relevant for $(g-2)_{\mu}$~\cite{Masjuan:2017tvw,Nyffeler:2016gnb}. While a wealth of experimental data is available in the single-virtual region [i.e. for $F_{P\gamma^*\gamma^*}(q^2,0)$] from $e^+e^-\to e^+e^-P$ collisions and $P\to\gamma \ell^+\ell^-$ decays, it is only in the past year that the first doubly-virtual measurement from $e^+e^-$ colliders for the $\eta'$ became available~\cite{BaBar:2018zpn}. Still, it would be desirable to gather complementary measurements at lower energies in order to test the most compelling frameworks devised for describing such TFFs in the region relevant to $(g-2)_{\mu}$~\cite{Masjuan:2017tvw,Hoferichter:2018kwz}, and also to compare against the recent lattice results~\cite{Gerardin:2019vio}. 
Experimentally, such a possibility can be brought by the double-Dalitz decays, $P\to\gamma \ell^+\ell^-\ell'^+\ell'^-$, especially for the $\eta$ and $\eta'$ cases due to their larger phase space~\cite{Bijnens:1999jp,Lih:2009np,Petri:2010ea,Terschlusen:2013iqa,DAmbrosio:2013qmd,Escribano:2015vjz,Weil:2017knt}.\footnote{Also, this could be feasible at low energies for the $\pi^0$ at BES~III~\cite{Redmer:2017fhg,Redmer:2018gah}, see comments in Section~\ref{sec:C&O} in this respect.} Notoriously, the required statistics for such measurements would be at hand at the proposed REDTOP facility~\cite{Gatto:2016rae}. Still, an accurate extraction of the TFF would not be possible without a proper implementation of QED radiative corrections. The first calculation of such kind for the double Dalitz decays appeared in \cite{Barker:2002ib}, that included all diagrams necessary to cancel IR divergencies, but did not represent a full NLO calculation, with some diagrams/topologies missing. In our work~\cite{Kampf:2018wau}, we revised such calculation, performing a full NLO evaluation in the soft-photon approximation. As a result, we find corrections to the previous estimate that are likely relevant for the extraction of the doubly-virtual TFF.

\section{Full NLO analysis}

At LO, the amplitude for double-Dalitz decays $(P\to \ell^+\ell^-\ell'^+\ell'^-)$ reads
  \begin{equation}
    i\mathcal{M}^{LO} = -ie^4\frac{F_{P\gamma\gamma}(s_{12},s_{34})}{s_{12} s_{34}}
                                 \epsilon_{\mu\nu\rho\sigma}p_{12}^{\mu}p_{34}^{\rho} 
                               \left[ \bar{u}(p_1)\gamma^{\nu}v(p_2)\right]  \left[\bar{u}(p_3)\gamma^{\sigma}v(p_4) \right], \label{eq:MLO}
  \end{equation}
with an additional exchange contribution ($2\leftrightarrow 3$) whenever identical leptons appear in the final state. For $\ell\neq\ell'$, this implies 
  \begin{equation}
    |\mathcal{M}^{LO}|^2 = \frac{e^8 |F_{P\gamma\gamma}(s_{12},s_{34})|^2}{x_{12}x_{34}}\lambda^2 
    \Big( 2 - \lambda_{12}^2 + y_{12}^2 -\lambda_{34}^2 + y_{34}^2 +(\lambda_{12}^2 - y_{12}^2)(\lambda_{34}^2 - y_{34}^2)\sin^2\phi\Big),
  \end{equation}
where the phase-space variables $\{s_{ij}, y_{ij}, \phi\}$, and also $\lambda_{(ij)}$, have been defined in \cite{Kampf:2018wau,Barker:2002ib}. Such result is modified at higher orders in QED. Particularly, denoting the generic amplitude as 
\begin{equation}
  |\mathcal{M}|^2 = |\mathcal{M}^{LO} +\mathcal{M}^{NLO} +...|^2 =  |\mathcal{M}^{LO}|^2 
                    +2\operatorname{Re}\mathcal{M}^{NLO}\mathcal{M}^{LO*} + ... = \textrm{LO} +\textrm{NLO},
\end{equation}
the NLO contribution to the differential branching ratio stands for the interference of the LO and NLO amplitudes above. 
\begin{figure}
  \includegraphics[width=0.9\textwidth]{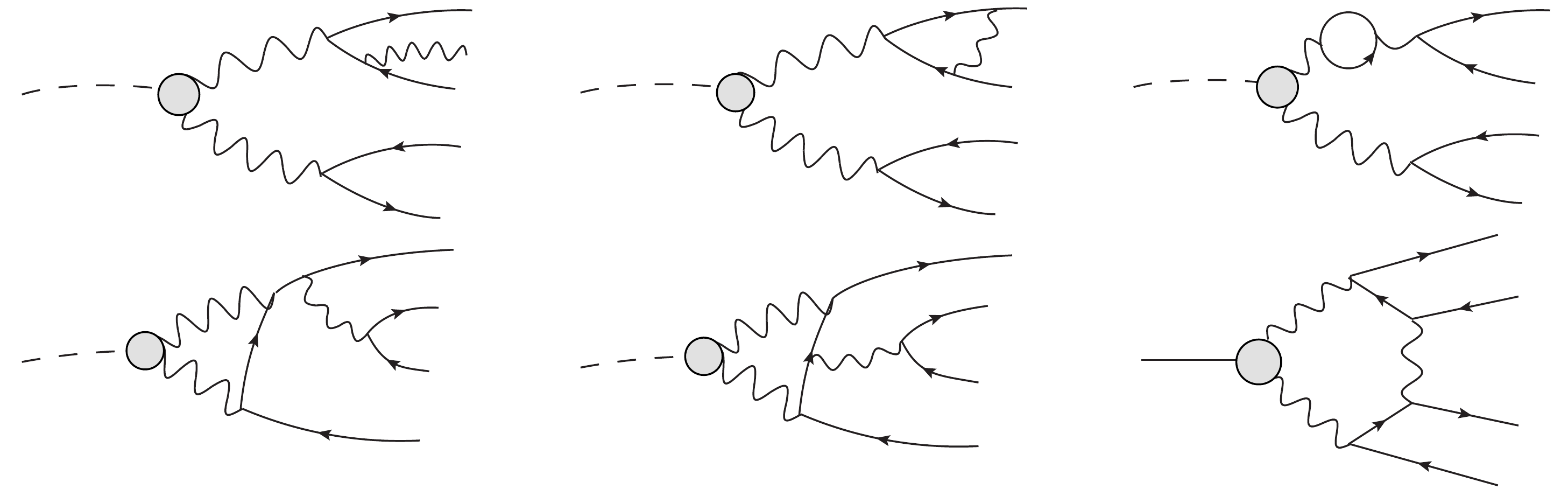}
  \caption{Representatives for each of the NLO contributions. From left to right, up to down: bremsstrahlung, vertex correction (self-energies are implicit), vacuum polarization, three-, four-, and five-point amplitudes. \label{fig:nlo}}
\end{figure}

Representative diagrams contributing to the NLO amplitudes are shown in Fig.~\ref{fig:nlo}. Among these diagrams, the vertex, vacuum polarization, and five-point amplitudes, all of them containing IR divergencies, where accounted for already in \cite{Barker:2002ib}, while the three- and four-point amplitudes where missing. The last types constitute, together with the revision of previous topologies, our main contribution. Further, in order to obtain a finite infrared (IR) result, bremsstrahlung diagrams need to be incorporated as well (see Fig.~\ref{fig:nlo}). Employing the soft-photon approximation, these can be expressed as 
\begin{equation}
  |\mathcal{M}^{BS}|^2 = -e^2 |\mathcal{M}^{LO}|^2 \sum_{i,j} Q_i Q_j \ I(p_i,p_j),
\end{equation}
where $Q_k$ and $p_k$ stand for the charge and momentum of the $k$-th particle respectively, and $I(p_i,p_j)$ are well-known integrals~\cite{tHooft:1978jhc} that can be found, adapted to our case of interest, in \cite{Kampf:2018wau}\footnote{We note that such integrals were not clearly defined for the $p_i=p_j$ case in \cite{Barker:2002ib}, which might contribute to the observed differences (see Sction~\ref{sec:num}).}. In the following, we revise each of the contributions as compared to Ref.~\cite{Barker:2002ib}. 

Concerning the vertex corrections, these amount to shifting the $\gamma^{\mu}$ Dirac matrices in Eq.~\ref{eq:MLO} to 
  \begin{equation}\label{eq:vertex}
    \gamma^{\mu} \to \gamma^{\mu}F_1(q^2) + \frac{i\sigma^{\mu\lambda}}{2m_{\ell}}q_{\lambda}F_2(q^2) 
  \end{equation}
where $F_1(q^2)$ and $F_2(q^2)$ are the well-known NLO contributions to the Dirac and Pauli form factors. The first one needs to be regularized, together with the self-energies, and contains IR divergencies to be compensated by bremsstrahlung graphs. Its inclusion amounts to a multiplicative factor with respect to the LO result and is in good agreement with Ref.~\cite{Barker:2002ib}. Concerning the second one, while we agree on the expression for $F_2$, we find differences with respect to the resulting matrix element squared. While its impact on the total branching ratio (BR) is negligible~\cite{Kampf:2018wau}, it might affect (softly) the differential distributions. Regarding the vacuum polarization, we find good agreement. 

Concerning the three- and four-point functions, these have been computed for the first time. In comparison to the vertex functions, these involve an integral over the TFF that, for computational purposes, has been assumed that can be taken as constant or be expressed in a propagator-like form (modulo powers of momenta). Further, for the sake of simplicity, a simple factorized Pad\'e approximant has been employed for numerical results. Analytic expressions for them have been provided, for $\ell\neq\ell'$, in \cite{Kampf:2018wau} in terms of three- and four-point tensor integrals---the full result has been made accessible as ancillary files on arXiv. 
The three-point topologies deserve further comments since, unlike for the four- and five-point ones, the amplitude itself is divergent for a constant TFF. In our work, two possibilities have been investigated: to use the form factor or to use chiral perturbation theory ($\chi$PT), where a counterterm that is common to $P\to\ell^+\ell^-$ decays is required to render the amplitude finite. For light pseudoscalars and leptons, the three-point topology is very similar in both approaches, while the performance of $\chi$PT (at this order) deteriorates for heavier pseudoscalars and leptons.

Finally, the five-point topology has been recomputed employing different techniques as compared to Ref.~\cite{Barker:2002ib}, that uses a tensor decomposition of the five-point loop integrals together with spinor amplitudes. In particular, in our final calculations, we compute directly the NLO piece, after summing over polarizations. This allows to decompose the integral in terms of the scalar five-point function and lower-point tensor ones. As a cross-check, we computed the NLO amplitude in terms of five-point tensor integrals that were decomposed into lower-point ones using the method in \cite{Denner:2002ii}. Such method overcomes some numerical instabilities encountered in Ref.~\cite{Barker:2002ib} connected to vanishing Gram determinants, and resulted in identical numerical results as the first method. Since such an integral contains IR divergencies, we also checked that, once combined with bremsstrahlung graphs, the result was IR-finite. Since final expressions are lengthy, it was not possible to compare to the previous results in \cite{Barker:2002ib} that might be a source of the mild discrepancies that we found.

\section{Numerical results}\label{sec:num}

In the following, we provide results on the correction to the total decay width with a soft-photon cut-off energy $E_c$, corresponding to $x_{4\ell}=0.9985$, in analogy to \cite{Barker:2002ib}.\footnote{$E_c =m_P(1-x_{4\ell})/2$ and $x_{4\ell} =p_{4\ell}^2 m_P^{-2}$. For relating to different $x_{4\ell}$ values, we refer to \cite{Kampf:2018wau}.} For the calculation, we employed LoopTools~\cite{Hahn:1998yk,Hahn:2006qw} to evaluate the loop integrals that we checked provided the same results as our implementation of the methods in \cite{Denner:2002ii}. For the numerical integrals we employed Mathematica's NIntegrate method as well as the MonteCarlo routines in \cite{Hahn:2004fe} for each of the contributions,\footnote{For details on results for each contribution separated, we refer to \cite{Kampf:2018wau}.} providing compatible results. As a cross-check, the total contribituon of the five-point amplitudes was compatible with $0$ within integration errors, providing an additional cross-check of the correctness of the code~\cite{Kampf:2018wau}. 
\begin{table}[!htb]
\centering
\resizebox{\textwidth}{!}{\small
  \begin{tabular}{cccccccc}\hline
                   & $\pi^0\to 4e$ & $K_L\to 4e$ & $K_L\to 2e2\mu$ & $K_L\to 4\mu$ & $\eta\to 4e$ & $\eta\to 2e2\mu$ & $\eta\to 4\mu$  \\ \hline
  $\delta(\textrm{NLO})$ &   $-0.1727(2)$ & $-0.2345(1)$ & $-0.0842(2)$ & $0.0608(2)$ & $-0.2409(1)$ & $-0.0900(1)$ & $0.0455(2)$ \\
  $\delta(\textrm{FF})$ &  $\phantom{-}0.0037(2)$ & $\phantom{-}0.0749(2)$ & $\phantom{-}0.6942(2)$ & $0.8608(3)$ & $\phantom{-}0.0207(2)$ & $\phantom{-}0.4829(2)$ & $0.6202(3)$ \\ \hline
  no 3,4       &$-0.1718(2)$ &$-0.2262(2)$ & $-0.0767(1)$ & $0.0704(1)$ & $-0.2301(1)$ & $-0.0836(1)$ & $0.0535(1)$ \\ 
  Barker              &$-0.160(2)$  &$-0.218(1)$  & $-0.066(1)$ & $0.084(1)$ & $-$ & $-$ & $-$ \\  \hline
  BR(LO+NLO)              &$2.840(1)10^{-5}$  &$5.120(1)10^{-5}$  & $4.436(1)10^{-6}$ & $1.851(1)10^{-9}$ & $5.202(1)10^{-5}$ & $5.393(1)10^{-6}$ & $10.289(2)10^{-9}$ \\  \hline
  \end{tabular}
}
\caption{Numerical results for the radiative corrections (second row). The third shows the effects of including a TFF on the LO result. The fourth row contains the same corrections included in Ref.~\cite{Barker:2002ib} (in fifth row), that show the disagreement. The last row stands for the BR at NLO accuracy.}
\label{tab:res}
\end{table}
The main results for the BR are given in Table~\ref{tab:res}, that displays the quantity $\delta(NLO) = \textrm{BR}^{NLO}/\textrm{BR}^{LO}-1$, where NLO superscripts refers to the results containing LO and NLO contributions to the BR. The second row, $\delta(\textrm{NLO})$, is our main result, obtained for a non-constant TFF, which effect is outlined in the third row. The fourth includes only those contributions in \cite{Barker:2002ib} and a constant TFF, so that it can be readily compared to~\cite{Barker:2002ib}, fifth row. 

As advanced, we find some discrepancies---of the order of $1\%$ compared to \cite{Barker:2002ib}. The source of this is unclear: a few possibilities have been outlined, but numerics in their MC generator could contribute too. Finally, we note that for heavier pseudoscalars the previously ommitted three- and four-point topologies represent an additional $1\%$ effect---likely relevant for extracting information about the doubly-virtual TFF~\cite{Escribano:2015vjz}, yet a quantitative statement would require a MC simulation.

\section{Connection with $e^+e^-\to e^+e^- P$ production}

\begin{figure}[t]
\centering
  \includegraphics[width=\textwidth]{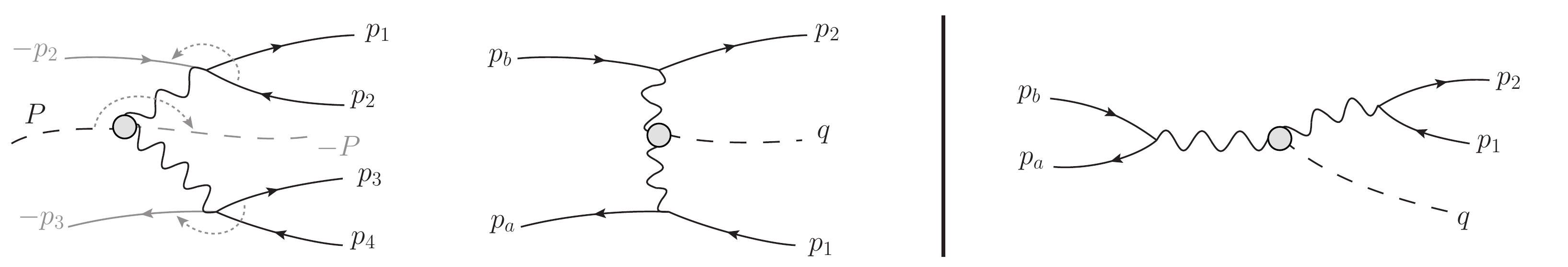}
\caption{Left block shows the choice for crossing relations from direct double-Dalitz terms (left) to the $e^+e^- \to e^+e^-P$ $t$-channel ones (center). The right block is the s-channel contribution to the latter process that would be crossed-related to the double-Dalitz exchange terms.\label{fig:crossing}}
  \end{figure}

As an extension, this process can be related via crossing symmetry to $e^+e^-\to e^+e^-P$ production~\cite{Sanchez-Puertas:2018zqc} (see Fig.~\ref{fig:crossing}), yet bremsstrahlung diagrams need to be recomputed in the appropriate reference frame---that of $e^+e^-$. As such, these results can be employed to complement the current NLO calculation in EKHARA~3.0~\cite{Czyz:2018jpp}, that misses the three-, four-, and five-point topologies and is a work in progress. 
For completeness, we detail here the required replacements relating both processes via crossing symmetry. With the choice in Fig.~\ref{fig:crossing}, the amplitudes referred in \cite{Kampf:2018wau} as Direct(Exchange) connect to the $t$- and $s$- channels for the mentioned process. Particularly, this is realized at the amplitude level through:
  \begin{align}
    p_1 \to p_2, && p_2 \to -p_b, && p_3 \to -p_a, && p_4 \to p_1, && P \to -q,  \nonumber\\
   \bar{u}_1 \to \bar{u}_2, && v_2 \to u_b\phantom{-}, && \bar{u}_3 \to \bar{v}_a\phantom{-}, && v_4\to v_1. && \phantom{P \to -q}
  \end{align}
Defining the following invariants
  \begin{equation}
    (p_a+p_b)^2 = s, \ (p_a-p_1)^2 = t_1, \ (p_b-p_2)^2 = t_2, \  (q+p_1)^2 = s_1, \  (q+p_2)^2 = s_2, 
  \end{equation}
it implies the following connection\footnote{For details on the phase space for $P\to \ell\bar{\ell}\ell\bar{\ell}$, we refer to \cite{Kampf:2018wau}, while for the $e^+e^- \to e^+e^-P$ process we employ that in Refs.~\cite{Schuler:1997ex,Czyz:2010sp}}
  \begin{align}
    2 p_{12}\cdot p_{34} &{}\to & 2(p_a-p_1)\cdot(p_b-p_2) &{}= M^2 -t_1 -t_2, \\ 
    2\bar{p}_{12}\cdot p_{34} &{}\to & -2(p_2+p_b)\cdot(p_a-p_1) &{}= 2m^2 +M^2 -2s_2 +t_1 -t_2,  \\
    2\bar{p}_{34}\cdot p_{12} &{}\to & 2(p_1+p_a)\cdot(p_b-p_2) &{}= -2m^2 -M^2 +2s_1 +t_1 -t_2,  \\
    2\bar{p}_{12}\cdot \bar{p}_{34} &{}\to & -2(p_1+p_a)\cdot(p_b+p_2) &{}= 4 m^2 -M^2 -4 s +2s_1 +2s_2 -t_1 -t_2,  \\
    (\epsilon^{p_1 p_2 p_3 p_4})^2 &{}\to & (\epsilon^{p_a p_b p_1 p_2})^2 &{}= -\Delta_4, 
  \end{align}
where $\Delta_4$ is the corresponding $4\times 4$ Gram determinant. This allows to define the crossed-related version of the old variables as 
  \begin{align}
    M^4\lambda^2 &{}\to & \bar{\lambda}^2 &{}= (M^2-t_1-t_2)^2 -4t_1t_2, \\
    y_{12} &{}\to & \bar{y}_{12} &{}= (2m^2 +M^2 -2s_2 +t_1 -t_2)\bar{\lambda}^{-1}, \\
    y_{34} &{}\to & \bar{y}_{34} &{}= (-2m^2 -M^2 +2s_1 +t_1 -t_2)\bar{\lambda}^{-1}, \\
    \cos{\phi} &{}\to & \cos{\bar{\phi}} &{}= \frac{ (M^2-t_1-t_2)\bar{y}_{12}\bar{y}_{34} - (4 m^2 -M^2 -4 s +2s_1 +2s_2 -t_1 -t_2)}
                                            {\left[4(t_2-4m^2-t_2 \bar{y}_{12})(t_1-4m^2-t_1 \bar{y}_{34})\right]^{1/2}} , \\
    \sin^2{\phi} &{}\to & \sin^2{\bar{\phi}} &{}= -16^2\Delta_4 \left(4(t_2-4m^2-t_2 \bar{y}_{12})(t_1-4m^2-t_1 \bar{y}_{34} \right)^{-1}. 
  \end{align}
%
Finally, the Bremsstrahlung contribution, with the soft-photon energy in the colliding $e^+e^-$ frame, reads
  \begin{align}
    e^2 I(p_i,p_j) = \frac{\alpha}{\pi} \frac{1+\beta_{ij}^2}{4\beta_{ij}} \Bigg[ 
                       2\ln\left( \frac{1+\beta_{ij}}{1-\beta_{ij}} \right) \ln\left( \frac{2E_c}{m_{\gamma}} \right) 
      + \frac{1}{4}\ln^2\left( \frac{\Omega_i^-}{\Omega_i^+} \right)  - \frac{1}{4}\ln^2\left( \frac{\Omega_j^-}{\Omega_j^+} \right) \nonumber\\ 
              +\operatorname{Li}_2\left( 1 - \frac{\Upsilon_{ij}\Omega_i^-}{\beta_{ij}} \right)
              +\operatorname{Li}_2\left( 1 - \frac{\Upsilon_{ij}\Omega_i^+}{\beta_{ij}} \right)
              -\operatorname{Li}_2\left( 1 - \frac{\Upsilon_{ij}\Omega_j^-}{\beta_{ij}\alpha_{ij}} \right)
              -\operatorname{Li}_2\left( 1 - \frac{\Upsilon_{ij}\Omega_j^+}{\beta_{ij}\alpha_{ij}} \right)
 \Bigg],	
  \end{align}
with $\beta_{ij}^2 = 1 -4m^2(p_i+p_j)^{-2}$ and $\alpha_{ij} = (1+\beta_{ij})/(1-\beta_{ij})^{-1}$, $\Upsilon_{ij} = 2(\alpha p_i^0 -p_j^0)(p_i+p_j)^{-2}$. The additional parameters are given by 
  \begin{align}
    (p_{a(b)} +p_{1(2)})^2 &{}= 4m^2 -t_{1(2)} \equiv s_{a1(b2)}, \\
    (p_{a(b)} +p_{2(1)})^2 &{}= m^2 +s +t_{2(1)} -s_{1(2)}  \equiv s_{a2(b1)}, \\
    (p_1 +p_2)^2           &{}= 2m^2+M^2+s-s_1-s_2,  \equiv s_{12} \\
    \Omega^{\pm}_{a,b}     &{}= \sqrt{s}(1\pm\beta)/2, \\	
    \Omega^{\pm}_{1(2)}    &{}= \left[ s+m^2-s_{2(1)} + \lambda^{1/2}(s,s_{2(1)},m^2) \right](2\sqrt{s})^{-1}, \\
    \Upsilon_{ab}          &{}= 2\beta(\sqrt{s}(1-\beta))^{-1}, \\
    \Upsilon_{a2(b1)}      &{}= \left[s(\alpha_{a2(b1)}-1) +s_1 -m^2\right](s_{a2(b1)}\sqrt{s})^{-1}, \\
    \Upsilon_{a1(b2)}      &{}= \left[s(\alpha_{a1(b2)}-1) +s_2 -m^2\right](s_{a1(b2)}\sqrt{s})^{-1}, \\
    \Upsilon_{12}          &{}= \left[(s+m^2)(\alpha_{12}-1) +s_1 -\alpha_{12}s_2\right](s_{12}\sqrt{s})^{-1}.
  \end{align}
with $s_{ij}\equiv (p_i+p_j)^2$ and $\lambda(a,b,c)\equiv a^2+b^2+c^2-2ab-2ac-2bc$. Finally, for $p_i=p_j$,
  \begin{align}
    e^2 I(p_{a,b},p_{a,b})   &{}= \frac{\alpha}{\pi} \left[ \ln\left( \frac{2E_c}{m_{\gamma}} \right) 
                                    +\frac{1}{2\beta}\ln\left( \frac{1-\beta}{1+\beta} \right) \right], \\
    e^2 I(p_{1(2)},p_{1(2)}) &{}= \frac{\alpha}{\pi} \left[ \ln\left( \frac{2E_c}{m_{\gamma}} \right) 
                      +\frac{s+m^2-s_{2(1)}}{2\lambda^{1/2}(s,s_{2(1)},m^2)}\ln\left( \frac{\Omega^-_{1(2)}}{\Omega^+_{1(2)}} \right) \right].
  \end{align}

\section{Conclusions and Outlook}\label{sec:C&O}

In this work, we have revisited and completed the full NLO calculation for double-Dalitz pseudoscalar decays in the soft-photon approximation. In doing so, we find small numerical differences with respect to the topologies already accounted for in the existing calculation, the source of which could not be determined precisely. In addition, the new topologies have a small but likely non-negligible effect for the extraction of the doubly-virtual TFF. The results have been provided in a Mathematica notebook available at arXiv as an ancillary file. Also, this work is relevant for completing the NLO corrections for pseudoscalar production at $e^+e^-$ colliders, which is currently a work in progress.



\acknowledgments 
This work was supported by the Czech Science Foundation (grant no. GACR 18-17224S), and by the project UNCE/SCI/013 of Charles University and also by the Ministerio de Ciencia, Innovaci{\'o}n y Universidades under the grant SEV-2016-0588, and the grant 754510 (EU, H2020-MSCA- COFUND-2016).

\bibliographystyle{JHEP}
\bibliography{mybib}

\providecommand{\href}[2]{#2}\begingroup\raggedright\begin{thebibliography}{10}

\bibitem{Lepage:1980fj}
G.~P. Lepage and S.~J. Brodsky, \emph{{Exclusive Processes in Perturbative
  Quantum Chromodynamics}},
  \href{https://doi.org/10.1103/PhysRevD.22.2157}{\emph{Phys. Rev.} {\bfseries
  D22} (1980) 2157}.

\bibitem{Lepage:1979zb}
G.~P. Lepage and S.~J. Brodsky, \emph{{Exclusive Processes in Quantum
  Chromodynamics: Evolution Equations for Hadronic Wave Functions and the
  Form-Factors of Mesons}},
  \href{https://doi.org/10.1016/0370-2693(79)90554-9}{\emph{Phys. Lett.}
  {\bfseries 87B} (1979) 359}.

\bibitem{Jegerlehner:2009ry}
F.~Jegerlehner and A.~Nyffeler, \emph{{The Muon g-2}},
  \href{https://doi.org/10.1016/j.physrep.2009.04.003}{\emph{Phys. Rept.}
  {\bfseries 477} (2009) 1} [\href{https://arxiv.org/abs/0902.3360}{{\ttfamily
  0902.3360}}].

\bibitem{Benayoun:2014tra}
T.~Blum, ed., \emph{{Hadronic contributions to the muon anomalous magnetic
  moment Workshop. $(g-2)_{\mu}$: Quo vadis? Workshop. Mini proceedings}},
  2014.

\bibitem{Masjuan:2017tvw}
P.~Masjuan and P.~Sanchez-Puertas, \emph{{Pseudoscalar-pole contribution to the
  $(g_{\mu}-2)$: a rational approach}},
  \href{https://doi.org/10.1103/PhysRevD.95.054026}{\emph{Phys. Rev.}
  {\bfseries D95} (2017) 054026}
  [\href{https://arxiv.org/abs/1701.05829}{{\ttfamily 1701.05829}}].

\bibitem{LeeRoberts:2011zz}
{\scshape Fermilab P989} collaboration, \emph{{The Fermilab muon (g-2)
  project}},
  \href{https://doi.org/10.1016/j.nuclphysbps.2011.06.038}{\emph{Nucl. Phys.
  Proc. Suppl.} {\bfseries 218} (2011) 237}.

\bibitem{Mibe:2010zz}
{\scshape J-PARC g-2} collaboration, \emph{{New g-2 experiment at J-PARC}},
  \href{https://doi.org/10.1088/1674-1137/34/6/022}{\emph{Chin. Phys.}
  {\bfseries C34} (2010) 745}.

\bibitem{Nyffeler:2016gnb}
A.~Nyffeler, \emph{{Precision of a data-driven estimate of hadronic
  light-by-light scattering in the muon $g-2$: Pseudoscalar-pole
  contribution}}, \href{https://doi.org/10.1103/PhysRevD.94.053006}{\emph{Phys.
  Rev.} {\bfseries D94} (2016) 053006}
  [\href{https://arxiv.org/abs/1602.03398}{{\ttfamily 1602.03398}}].

\bibitem{BaBar:2018zpn}
{\scshape BaBar} collaboration, \emph{{Measurement of the
  $\gamma^{\star}\gamma^{\star} \to \eta'$ transition form factor}},
  \href{https://doi.org/10.1103/PhysRevD.98.112002}{\emph{Phys. Rev.}
  {\bfseries D98} (2018) 112002}
  [\href{https://arxiv.org/abs/1808.08038}{{\ttfamily 1808.08038}}].

\bibitem{Hoferichter:2018kwz}
M.~Hoferichter, B.-L. Hoid, B.~Kubis, S.~Leupold and S.~P. Schneider,
  \emph{{Dispersion relation for hadronic light-by-light scattering: pion
  pole}}, \href{https://doi.org/10.1007/JHEP10(2018)141}{\emph{JHEP} {\bfseries
  10} (2018) 141} [\href{https://arxiv.org/abs/1808.04823}{{\ttfamily
  1808.04823}}].

\bibitem{Gerardin:2019vio}
A.~G{\'e}rardin, H.~B. Meyer and A.~Nyffeler, \emph{{Lattice calculation of the
  pion transition form factor with $N_f=2+1$ Wilson quarks}},
  \href{https://arxiv.org/abs/1903.09471}{{\ttfamily 1903.09471}}.

\bibitem{Bijnens:1999jp}
F.~Perrsson, \emph{{Effects of different form-factors in meson photon photon
  transitions and the muon anomalous magnetic moment}},  Master's thesis, Lund
  U., Dept. Theor. Phys., 1999.

\bibitem{Lih:2009np}
C.-C. Lih, \emph{{Study of pi0 and eta decays containing dilepton}},
  \href{https://doi.org/10.1088/0954-3899/38/6/065001}{\emph{J. Phys.}
  {\bfseries G38} (2011) 065001}
  [\href{https://arxiv.org/abs/0912.2147}{{\ttfamily 0912.2147}}].

\bibitem{Petri:2010ea}
T.~Petri, \emph{{Anomalous decays of pseudoscalar mesons}}, Ph.D. thesis,
  Julich, Forschungszentrum, 2010.
\newblock \href{https://arxiv.org/abs/1010.2378}{{\ttfamily 1010.2378}}.

\bibitem{Terschlusen:2013iqa}
C.~Terschl{\"u}sen, B.~Strandberg, S.~Leupold and F.~Eichstädt,
  \emph{{Reactions with pions and vector mesons in the sector of odd intrinsic
  parity}}, \href{https://doi.org/10.1140/epja/i2013-13116-6}{\emph{Eur. Phys.
  J.} {\bfseries A49} (2013) 116}
  [\href{https://arxiv.org/abs/1305.1181}{{\ttfamily 1305.1181}}].

\bibitem{DAmbrosio:2013qmd}
G.~D'Ambrosio, D.~Greynat and G.~Vulvert, \emph{{Standard Model and New Physics
  contributions to $K_L$ and $K_S$ into four leptons}},
  \href{https://doi.org/10.1140/epjc/s10052-013-2678-1}{\emph{Eur. Phys. J.}
  {\bfseries C73} (2013) 2678}
  [\href{https://arxiv.org/abs/1309.5736}{{\ttfamily 1309.5736}}].

\bibitem{Escribano:2015vjz}
R.~Escribano and S.~Gonz{\`a}lez-Sol{\'i}s, \emph{{A data-driven approach to
  $\pi^{0}, \eta$ and $\eta^{\prime}$ single and double Dalitz decays}},
  \href{https://doi.org/10.1088/1674-1137/42/2/023109}{\emph{Chin. Phys.}
  {\bfseries C42} (2018) 023109}
  [\href{https://arxiv.org/abs/1511.04916}{{\ttfamily 1511.04916}}].

\bibitem{Weil:2017knt}
E.~Weil, G.~Eichmann, C.~S. Fischer and R.~Williams, \emph{{Electromagnetic
  decays of the neutral pion}},
  \href{https://doi.org/10.1103/PhysRevD.96.014021}{\emph{Phys. Rev.}
  {\bfseries D96} (2017) 014021}
  [\href{https://arxiv.org/abs/1704.06046}{{\ttfamily 1704.06046}}].

\bibitem{Redmer:2017fhg}
{\scshape BESIII} collaboration, \emph{{The two-photon physics program at
  BESIII}},
  \href{https://doi.org/10.1016/j.nuclphysbps.2017.03.053}{\emph{Nucl. Part.
  Phys. Proc.} {\bfseries 287-288} (2017) 99}.

\bibitem{Redmer:2018gah}
{\scshape BESIII} collaboration, \emph{{The $\gamma\gamma$ Physics Program at
  BESIII}}, \href{https://doi.org/10.1051/epjconf/201816600017}{\emph{EPJ Web
  Conf.} {\bfseries 166} (2018) 00017}.

\bibitem{Gatto:2016rae}
{\scshape REDTOP} collaboration, \emph{{The REDTOP project: Rare Eta Decays
  with a TPC for Optical Photons}},
  \href{https://doi.org/10.22323/1.282.0812}{\emph{PoS} {\bfseries ICHEP2016}
  (2016) 812}.

\bibitem{Barker:2002ib}
A.~R. Barker, H.~Huang, P.~A. Toale and J.~Engle, \emph{{Radiative corrections
  to double Dalitz decays: Effects on invariant mass distributions and angular
  correlations}}, \href{https://doi.org/10.1103/PhysRevD.67.033008}{\emph{Phys.
  Rev.} {\bfseries D67} (2003) 033008}
  [\href{https://arxiv.org/abs/hep-ph/0210174}{{\ttfamily hep-ph/0210174}}].

\bibitem{Kampf:2018wau}
K.~Kampf, J.~Novotn\'y and P.~Sanchez-Puertas, \emph{{Radiative corrections to
  double-Dalitz decays revisited}},
  \href{https://doi.org/10.1103/PhysRevD.97.056010}{\emph{Phys. Rev.}
  {\bfseries D97} (2018) 056010}
  [\href{https://arxiv.org/abs/1801.06067}{{\ttfamily 1801.06067}}].

\bibitem{tHooft:1978jhc}
G.~'t~Hooft and M.~J.~G. Veltman, \emph{{Scalar One Loop Integrals}},
  \href{https://doi.org/10.1016/0550-3213(79)90605-9}{\emph{Nucl. Phys.}
  {\bfseries B153} (1979) 365}.

\bibitem{Denner:2002ii}
A.~Denner and S.~Dittmaier, \emph{{Reduction of one loop tensor five point
  integrals}}, \href{https://doi.org/10.1016/S0550-3213(03)00184-6}{\emph{Nucl.
  Phys.} {\bfseries B658} (2003) 175}
  [\href{https://arxiv.org/abs/hep-ph/0212259}{{\ttfamily hep-ph/0212259}}].

\bibitem{Hahn:1998yk}
T.~Hahn and M.~Perez-Victoria, \emph{{Automatized one loop calculations in
  four-dimensions and D-dimensions}},
  \href{https://doi.org/10.1016/S0010-4655(98)00173-8}{\emph{Comput. Phys.
  Commun.} {\bfseries 118} (1999) 153}
  [\href{https://arxiv.org/abs/hep-ph/9807565}{{\ttfamily hep-ph/9807565}}].

\bibitem{Hahn:2006qw}
T.~Hahn and M.~Rauch, \emph{{News from FormCalc and LoopTools}},
  \href{https://doi.org/10.1016/j.nuclphysbps.2006.03.026}{\emph{Nucl. Phys.
  Proc. Suppl.} {\bfseries 157} (2006) 236}
  [\href{https://arxiv.org/abs/hep-ph/0601248}{{\ttfamily hep-ph/0601248}}].

\bibitem{Hahn:2004fe}
T.~Hahn, \emph{{CUBA: A Library for multidimensional numerical integration}},
  \href{https://doi.org/10.1016/j.cpc.2005.01.010}{\emph{Comput. Phys. Commun.}
  {\bfseries 168} (2005) 78}
  [\href{https://arxiv.org/abs/hep-ph/0404043}{{\ttfamily hep-ph/0404043}}].

\bibitem{Sanchez-Puertas:2018zqc}
P.~Sanchez-Puertas, K.~Kampf and J.~Novotn{\'y}, \emph{{A revision of radiative
  corrections to double-Dalitz decays}},
  \href{https://doi.org/10.1051/epjconf/201919902014}{\emph{EPJ Web Conf.}
  {\bfseries 199} (2019) 02014}
  [\href{https://arxiv.org/abs/1809.01868}{{\ttfamily 1809.01868}}].

\bibitem{Czyz:2018jpp}
H.~Czyz and P.~Kisza, \emph{{EKHARA 3.0: an update of the EKHARA Monte Carlo
  event generator}},
  \href{https://doi.org/10.1016/j.cpc.2018.07.021}{\emph{Comput. Phys. Commun.}
  {\bfseries 234} (2019) 245}
  [\href{https://arxiv.org/abs/1805.07756}{{\ttfamily 1805.07756}}].

\bibitem{Schuler:1997ex}
G.~A. Schuler, \emph{{Two photon physics with GALUGA 2.0}},
  \href{https://doi.org/10.1016/S0010-4655(97)00127-6}{\emph{Comput. Phys.
  Commun.} {\bfseries 108} (1998) 279}
  [\href{https://arxiv.org/abs/hep-ph/9710506}{{\ttfamily hep-ph/9710506}}].

\bibitem{Czyz:2010sp}
H.~Czyz and S.~Ivashyn, \emph{{EKHARA: A Monte Carlo generator for e+ e- to e+
  e- pi0 and e+ e- to e+ e- pi+ pi- processes}},
  \href{https://doi.org/10.1016/j.cpc.2011.01.029}{\emph{Comput. Phys. Commun.}
  {\bfseries 182} (2011) 1338}
  [\href{https://arxiv.org/abs/1009.1881}{{\ttfamily 1009.1881}}].

\end{thebibliography}\endgroup

%

\end{document}